\documentclass[epj]{svjour}

\def\be{\begin{equation}} 
\def\ee{\end{equation}}   

\RequirePackage{graphicx}

\usepackage{cite} 
\usepackage{amsmath}
\usepackage{hyperref}
\usepackage{cite}
\usepackage{amsmath,amssymb}
\usepackage{xcolor}
\usepackage[utf8]{inputenc}

\usepackage{graphics}

\usepackage{cite} 
\usepackage{amsmath}
\usepackage{hyperref}
\usepackage{cite}
\usepackage{amsmath,amssymb}
\usepackage{xcolor}

\begin{document}
\title{Interior solutions of relativistic stars in the scale-dependent scenario} 


\author{
Grigoris Panotopoulos \inst{1} 
\thanks{E-mail: \href{mailto:grigorios.panotopoulos@tecnico.ulisboa.pt}{\nolinkurl{grigorios.panotopoulos@tecnico.ulisboa.pt}} }
\and
\'Angel Rinc\'on \inst{2}
\thanks{E-mail: \href{mailto:angel.rincon@pucv.cl}{\nolinkurl{angel.rincon@pucv.cl}} }
\and
Il{\'i}dio Lopes \inst{1} 
\thanks{E-mail: \href{mailto:ilidio.lopes@tecnico.ulisboa.pt}{\nolinkurl{ilidio.lopes@tecnico.ulisboa.pt}} }
}
%
%
\institute{ 
Centro de Astrof\'{\i}sica e Gravita{\c c}{\~a}o, Departamento de F{\'i}sica, Instituto Superior T\'ecnico-IST, \\
Universidade de Lisboa-UL, Av. Rovisco Pais, 1049-001 Lisboa, Portugal
\and
Instituto de F{\'i}sica, Pontificia Universidad Cat{\'o}lica de Valpara{\'i}so, Avenida Brasil 2950, Casilla 4059, Valpara{\'i}so, Chile.
}
\date{Received: date / Revised version: date}
%
\abstract{
We study relativistic stars in the scale-dependent scenario, which is one of the approaches to quantum gravity, and where Newton's constant is promoted to a scale-dependent quantity. First, the generalized structure equations are derived here for the first time. Then they are integrated numerically assuming a linear equation-of-state in the simplest MIT bag model for quark matter. We compute the radius, the mass and the compactness of strange quarks stars, and we show that the energy conditions are fulfilled.
\PACS{
      {PACS-key}{discribing text of that key}   \and
      {PACS-key}{discribing text of that key}
     } 
} 
\maketitle

\section{Introduction}\label{Intro}

Einstein's General Relativity (GR) \cite{GR} is widely accepted as a relativistic theory of gravitation, which is at the same time both beautiful and very successful \cite{tests1,tests2}. Remarkably, a series of predictions have been confirmed observationally over the past 100 years or so. Apart from the classical tests and solar system tests \cite{tests3}, over the last four years or so two more predictions of GR, namely the existence of gravitational waves as well as the existence of black holes, have been confirmed as well thanks to the efforts of the LIGO/VIRGO collaborations \cite{Ligo}.

Despite its success, it has been known for a long time now that GR is not compatible with quantum physics. It is well-known that the formulation of a consistent quantum theory of gravity is still an open task in modern theoretical physics. Although as of today several approaches to the problem do exist in the literature (for a partial list see e.g. \cite{QG1,QG2,QG3,QG4,QG5,QG6,QG7,QG8,QG9} and references therein), there is one property in particular that all of those have in common. Namely, the basic quantities that enter into the action defining the model at hand, such as Newton's constant, electromagnetic coupling, the cosmological constant etc, become scale dependent (SD) quantities. This does not come as a surprise of course, since scale dependence at the level of the effective action is a generic feature of ordinary quantum field theory. 

In black hole physics the impact of the SD scenario on properties of black holes has been studied over the last years, and it has been found that the scale dependence modifies the horizon, the thermodynamics as well as the quasinormal spectra of classical black hole backgrounds \cite{SD1,SD2,SD3,SD4,SD5,SD6,SD7,ourprd}. However, the astrophysical implications of the SD scenario should be investigated as well. In the present work we propose to obtain for the first time interior solutions of relativistic stars in the SD scenario, see however \cite{alfio} for an alternative approach. In particular, here we shall focus on strange quark stars, which may be viewed as an alternative to the more conventional paradigm based on neutron stars. Although as of today it is a  speculative class of compact objects, strange quarks stars cannot conclusively be ruled out yet. As a matter of fact, there are some claims in the literature that there are currently some observed compact objects with peculiar features (such as small radii for instance) that cannot be explained assuming the known hadronic equations-of-state for neutron stars, see e.g. \cite{cand1,cand2,cand3}, and also Table 5 of \cite{weber} and references therein.

Our work is organized as follows: After this introduction, in the next section we briefly review the formalism of the SD scenario. In the third section we apply it to the case of relativistic stars, and we present and discuss our main numerical results. In the last section we close our work with some concluding remarks. We adopt the mostly positive metric signature, $(-,+,+,+)$, and we work in geometrical units where the speed of light in vacuum as well as the classical Newton's constant are set to unity, $c=1=G_N$.

\section{Scale-dependent gravity}

This section is devoted to summarize the main features about the formalism used along this manuscript, i.e., scale-dependent gravity. The idea is largely inspired by the asymptotic safety program and related approaches as the well-known Renormalization group improvement method \cite{Bonanno:2000ep,Bagnuls:2000ae,Bonanno:2001xi,Reuter:2003ca}. Taking them as inspiration, some authors introduced the now known scale-dependent gravity which has been systematically used in black holes physics
\cite{Koch:2016uso,Rincon:2017ypd,Rincon:2017goj,Rincon:2017ayr,
Contreras:2017eza,Rincon:2018sgd,Contreras:2018dhs,Rincon:2018lyd,
Rincon:2018dsq,Contreras:2018gct,Rincon:2019cix,Rincon:2019zxk,
Contreras:2019fwu,Fathi:2019jid,Panotopoulos:2019qjk,Contreras:2019nih}, and recently, in cosmological problems \cite{Hernandez-Arboleda:2018qdo,Canales:2018tbn}. The crucial point of scale-dependent gravity is to promote the classical parameter $\{A_0, B_0, (\cdots)_0\}$
to functions which depends on the energy scale $k$, namely $\{A_k, B_k, (\cdots)_k\}$. Depending of the case, the arbitrary scale $k$ is linked to the radial coordinate (assuming circular symmetry) or it is connected to the physical time.
In what follows, we will consider three different contributions accounted into the following action:
\begin{align} \label{action}
S[g_{\mu \nu},k] \equiv S_{\text{EH}} + S_{\text{M}} + S_{\text{SD}},
\end{align} 
where the first term, $S_{\text{EH}}$, is given by the well-known Einstein Hilbert action, the second term, $S_{\text{M}}$, is associated with a perfect fluid distribution, and the third term,  $S_{\text{SD}}$, encodes the scale-dependent sector.
In light of this, the scale-dependent coupling of the theory is the Newton’s coupling $G_k$ only (which can be related with the
gravitational coupling by $\kappa_k \equiv 8 \pi G_k$). In addition, there are two independent fields, which are the metric tensor $g_{\mu \nu}(x)$ and the scale field $k(x)$. 
The equations of motion obtained from a variation of \eqref{action} with respect to $g_{\mu \nu}(x)$ to obtain:
\begin{align}
G_{\mu \nu} \equiv R_{\mu \nu} - \frac{1}{2}R g_{\mu \nu} = \kappa_k T_{\mu \nu}^{\text{effec}},
\end{align}
where the effective energy momentum tensor, $T_{\mu \nu}^{\text{effec}}$, is defined in such a way that it include both the usual matter fields $T_{\mu \nu}$ and the contribution of the G-varying part $\Delta t_{\mu \nu}$. Then, such a tensor is defined to be
\begin{align}\label{Teffe}
\kappa_k T_{\mu \nu}^{\text{effec}} \equiv \kappa_k T_{\mu \nu} - \Delta t_{\mu \nu}.
\end{align}
Also, notice that the two tensor defined in the right hand side are given by the expressions:
\begin{equation}
T_{\nu}^{\mu} = \text{diag}(-\rho, p, p, p),
\end{equation}
and
\begin{equation}
\Delta t_{\mu \nu} = G_k \Bigl(g_{\mu \nu} \Box - \nabla_\mu \nabla_\nu\Bigl) G_k^{-1}
\end{equation}
At this level, some comments are in order. Firstly notice that, in a quantum field theory, the corresponding renormalization scale $k$ needs to be linked with the parameters of the physical system under study. 
For background solutions of the gap equations,
it is not constant any more. 
Therefore, an arbitrarily non-constant $k = k(x)$ implies that the
set of equations of motion does not close consistently.
The above means that the stress energy tensor is most likely
not conserved for almost any choice of the functional
dependence $k = k(x)$. This pathology has been considerably investigated 
in the context of renormalization group improvement of black holes in asymptotic safety scenarios.
To clarify the situation we should remark that the problem with the conservation law appears due one consistency equation is missing. The latter equation can be computed taking the variation of the corresponding action respect to the arbitrary scalar field $k(r)$, namely
\begin{align} \label{Seffec}
\frac{\mathrm{d}}{\mathrm{d}k}S[g_{\mu \nu},k]  &= 0,
\end{align}
which is usually interpreted as a variational scale setting procedure \cite{Reuter:2003ca,Koch:2010nn}.
To guarantee the adequate conservation of the stress energy tensor, we should to combine \eqref{Seffec} and the equations of motion. This, route, however, introduce a unavoidable problem, i.e., it is required first to compute the $\beta$-functions of the problem. It is well-known that the $\beta$-functions are not unique, reason why use them introduce some uncertainties. A way to circumvent the aforementioned issue is to impose some supplementary condition. 
In exterior solutions, the null energy condition (one of the four energy conditions demanded in general relativity) has been typically used as a supplementary equation. Thus, we promote the classical couplings to radial-dependent couplings (given that $\mathcal{O}(k(r)) \rightarrow \mathcal{O}(r)$) and, supported by the NEC, we are able to solve the functions involved. 
Thus, this philosophy of assuring the consistency of the equations by imposing a null energy condition will also be applied for the first time in the following study on interior solutions of relativistic stars.

\section{Hydrostatic equilibrium of relativistic stars}

Here we briefly review relativistic stars in GR, and then we generalize the structure equations in the SD scenario.

\subsection{Structure equations}

For non-rotating objects we seek spherically symmetric solutions in Schwarzschild coordinates, $(t,r,\theta,\phi)$, making as usual for the metric tensor the following ansatz
\begin{equation}
ds^2 = -e^{2 \nu} dt^2 + A(r) dr^2 + r^2 \mathrm{d \Omega^2},
\end{equation}
where $\mathrm{d \Omega^2} = d \mathrm{\theta^2} + \mathrm{sin^2 \theta \: d \phi^2}$ is the usual line element of the unit two-dimensional sphere. In addition, we introduce for convenience the mass function $m(r)$, which is defined to be
\begin{equation}
A(r)^{-1} \equiv 1 - \frac{2 m(r)}{r}
\end{equation}
Assuming for matter a fluid characterized by a stress-energy tensor of the form
\begin{equation}
T_\nu ^\mu = \text{diag}(-\rho, p, p, p)
\end{equation}
with $\rho$ being the energy density and $p$ being the pressure, the tt and rr field equations yield
\begin{eqnarray}
m'(r) & = & 4 \pi r^2 \rho(r) \\
\nu'(r) & = & \frac{m(r)+4 \pi r^3 p(r)}{r^2 (1-2m(r)/r)}
\end{eqnarray}
respectively, where a prime denotes differentiation with respect to the radial coordinate $r$. Instead of the angular field equations one may employ the conservation of energy
\begin{equation}
p'(r) = - [\rho(r) + p(r)] \nu'(r)
\end{equation}
Therefore, the Tolman-Oppenheimer-Volkoff equations \cite{Tolman,OV} read
\begin{eqnarray}
m'(r) & = & 4 \pi r^2 \rho(r) \\
p'(r) & = & - [\rho(r) + p(r)] \: \frac{m(r)+4 \pi r^3 p(r)}{r^2 (1-2m(r)/r)} \\
\nu'(r) & = & - \frac{p'(r)}{\rho(r)+p(r)}
\end{eqnarray}
supplemented by the appropriate equation-of-state $p(\rho)$ or $\rho(p)$. The first two equations are integrated imposing the initial conditions at the centre of the star, $m(0)=0, p(0)=p_c$.
As the interior solution and the exterior Schwarzschild vacuum solution \cite{SBH} are required to match at the surface of the star, the conditions $p(R)=0$ and $m(R)=M$ are used to compute the radius and the mass of the object. Finally, if desired, the metric potential $\nu$ can be also determined integrating the last equation and imposing the condition at the surface of the star
\begin{equation}
e^{2 \nu(R)} = 1 - \frac{2M}{R}
\end{equation}
Therefore, the solution for $\nu(r)$ is given by
\begin{equation}
\nu(r) = \nu(R) - \int_R^r \frac{p'(z)}{p(z)+\rho(z)}
\end{equation}
where 
\begin{align}
\nu(R)=\frac{1}{2} \ln\left(1-\frac{2M}{R}\right).
\end{align}
In the following step, we now wish to generalize the standard structure equations valid in GR in the SD approach to quantum gravity. To that end, we first recall that in this framework Newton's constant is promoted to a function of the radial coordinate, $G(r)$, and that in the field equations
\begin{align}
G_{\mu \nu} \equiv R_{\mu \nu} - \frac{1}{2}R g_{\mu \nu} = 8 \pi G(r) T_{\mu \nu}^{\text{effec}},
\end{align}
the total (or effective) stress-energy tensor has two contributions, namely one from the ordinary matter, $T_{\mu \nu}$, and another from the contribution of the G-varying part, $\Delta t_{\mu \nu}$
\begin{equation}
T_{\mu \nu}^{\text{effec}} = T_{\mu \nu} - \frac{1}{8 \pi G} \: \Delta t_{\mu \nu}
\end{equation}
where the G-varying part is given by \cite{formalism}
\begin{equation}
\Delta t_{\mu \nu} = G(r) \: (g_{\mu \nu} \Box - \nabla_\mu \nabla_\nu) \: G(r)^{-1}
\end{equation}
Therefore, it is straightforward to obtain the new structure equations, which are found to be
\begin{eqnarray}
(G(r) m(r))' & = & 4 \pi G(r) r^2 \rho_{eff}(r) \\
\nu'(r) & = & G(r) \: \frac{m(r)+4 \pi r^3 p_{eff}(r)}{r^2 (1-2G(r)m(r)/r)}
\end{eqnarray}
and we spare the details for the last equation, as it is too long to be included here. We have checked that the new generalized structure equations are reduced to the usual TOV equations when Newton's constant is taken to be a constant, $G'(r)=0=G''(r)$. Finally, there is an additional differential equation of second order for $G(r)$, which is the following \cite{formalism}
\begin{equation}
2\frac{G(r)''}{G(r)'} - 4 \frac{G(r)'}{G(r)} = \Bigl(\ln(\text{e}^{2\nu(r)}A(r)) \Bigl)'
\end{equation}
and which must be supplemented by two initial conditions at the centre of the star, $G(0)=G_c$ and $G'(0)=G_1$. In this framework the metric potential $A(r)$ and the mass function $m(r)$ are related via
\begin{equation}
A(r)^{-1} \equiv 1 - \frac{2 G(r) m(r)}{r}
\end{equation}
It should be pointed out that the exterior solution remains the same in the SD scenario assuming a vanishing cosmological constant. Therefore, the interior solution obtained assuming a r-varying gravitational constant will be matched to the usual Schwarzschild vacuum solution at the surface of the stars.

\subsection{Equation-of-state}

In order to close the system of equations, an equation-of-state for quark matter must be assumed. Here we shall adopt the very popular and widely used in the literature ``radiation plus constant" analytic function
\begin{equation}
p = \frac{1}{3} (\rho - 4B)
\end{equation}
which is the simplest version of the MIT bag model \cite{bagmodel1,bagmodel2}. The model is characterized by three parameters, namely the mass of the $s$ quark, $m_s$, the QCD coupling constant, $\alpha_c$, and the bag constant, $B$. In the simplest MIT bag model $m_s=0=\alpha_c$, while $B=60~MeV/fm^3$ \cite{farhi,simpleMIT}.

More realistic and sophisticated EoSs have been developed and considered over the years in the literature. For instance, at asymptotically large densities color superconductivity effects~\cite{wilczek1,wilczek2} become important. Quark matter is in the color flavor 
locked (CFL) state~\cite{wilczek3,wilczek4}, in which quarks form Cooper pairs of different color and flavor, and where all quarks have the same Fermi momentum and electrons cannot be present.
Other possibilities include models that incorporate a chiral symmetry breaking \cite{NJL1,NJL2}, or models based on perturbative QCD studies \cite{refine1,art}, and others \cite{refine2,refine3}.


\begin{table*}
\centering
\caption{Basic properties of the five interior solutions obtained here.}
\begin{tabular}{cccccc}
\hline
$No$ of solution &  $R[km]$ & $M[M_{\odot}]$ & $C=M/R$ & $G_c$ & $G_1[km^{-1}]$  \\
\hline
  GR   & 11.139 & 1.707  & 0.228 &  1         &   0         \\ 
\hline
1      & 10.971 & 1.548  & 0.210 &  1.00327   &  -0.00025   \\
\hline
2      & 9.286  & 0.794  & 0.127 &  0.99750   &  +0.00025   \\
\hline
3      & 9.250  & 0.800  & 0.129 &  1.00250   &  -0.00025   \\
\hline
4      & 10.962 & 1.508  & 0.205 &  0.99673   &  +0.00025   \\
\hline
\end{tabular}
\label{table:First_set}
\end{table*}


\subsection{Numerical results}


\begin{figure*}[ht]
\centering
\includegraphics[width=0.32\textwidth]{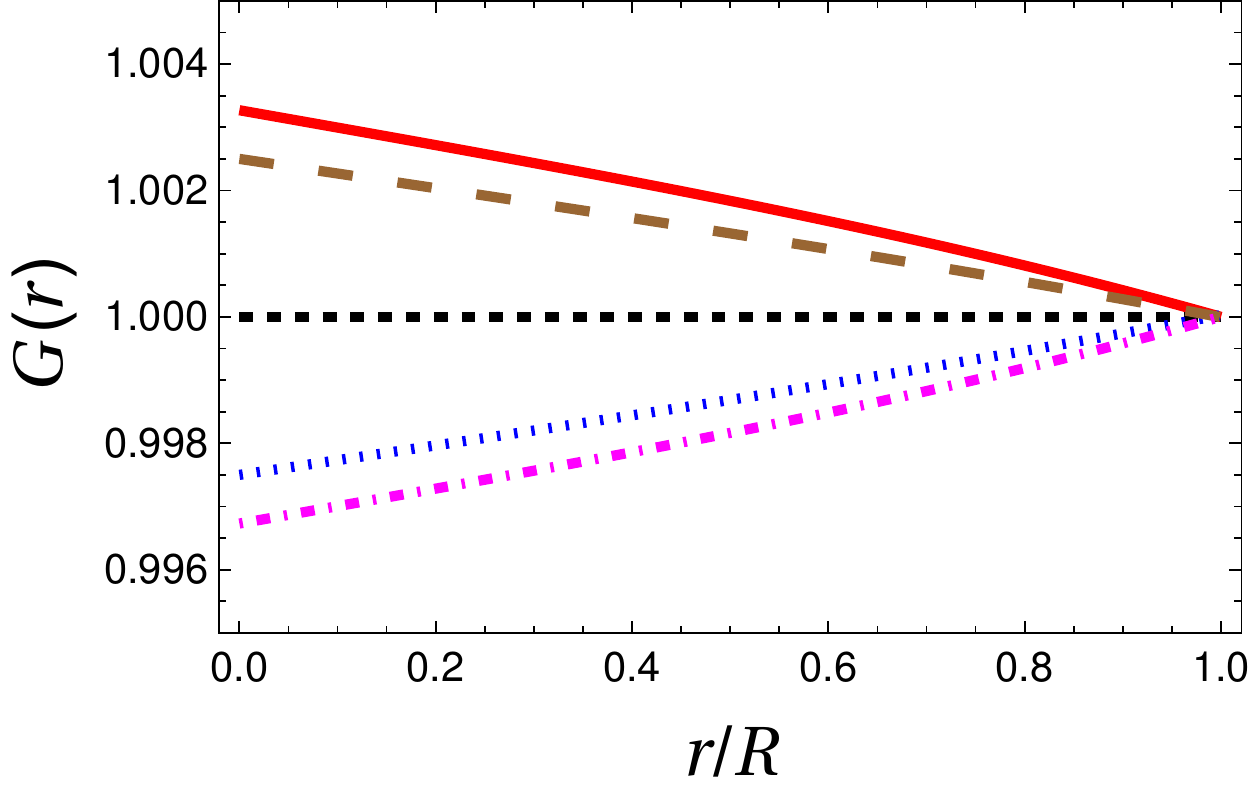}      \
\includegraphics[width=0.32\textwidth]{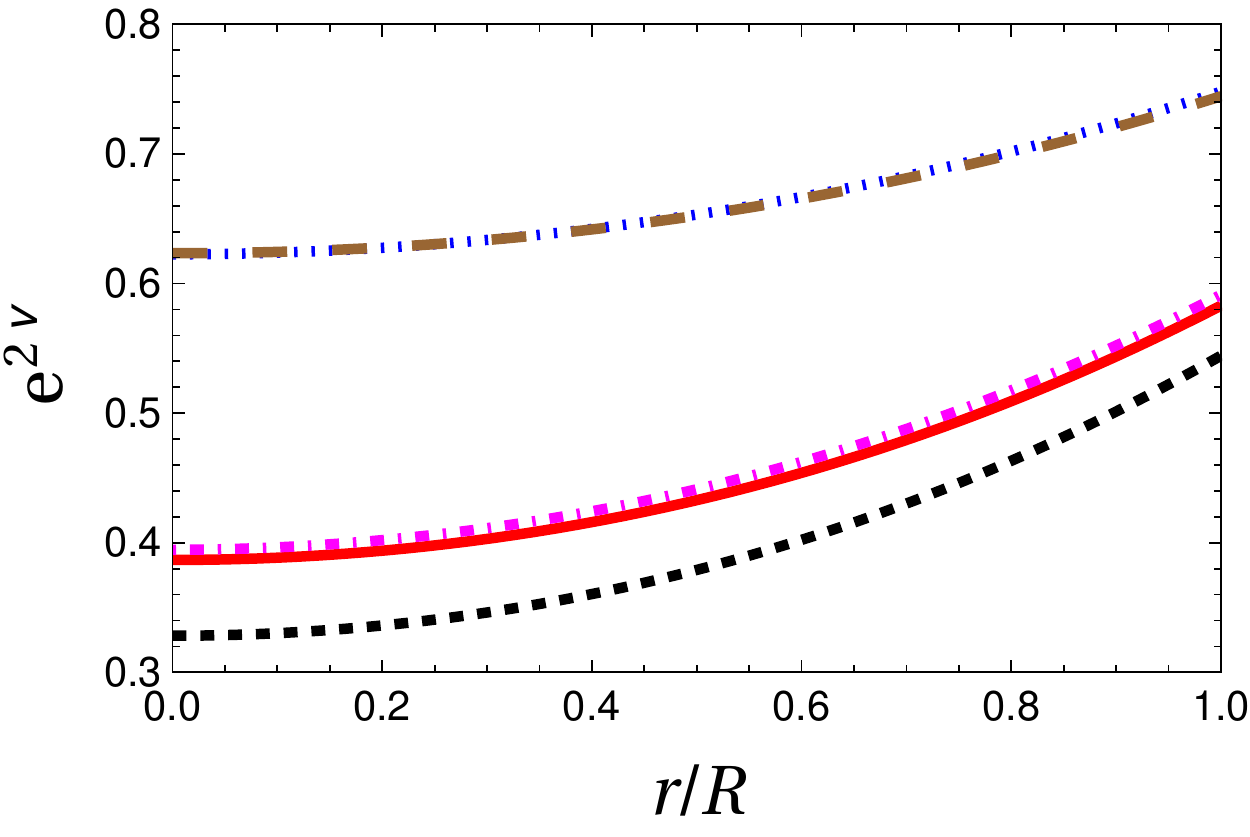}   \
\includegraphics[width=0.32\textwidth]{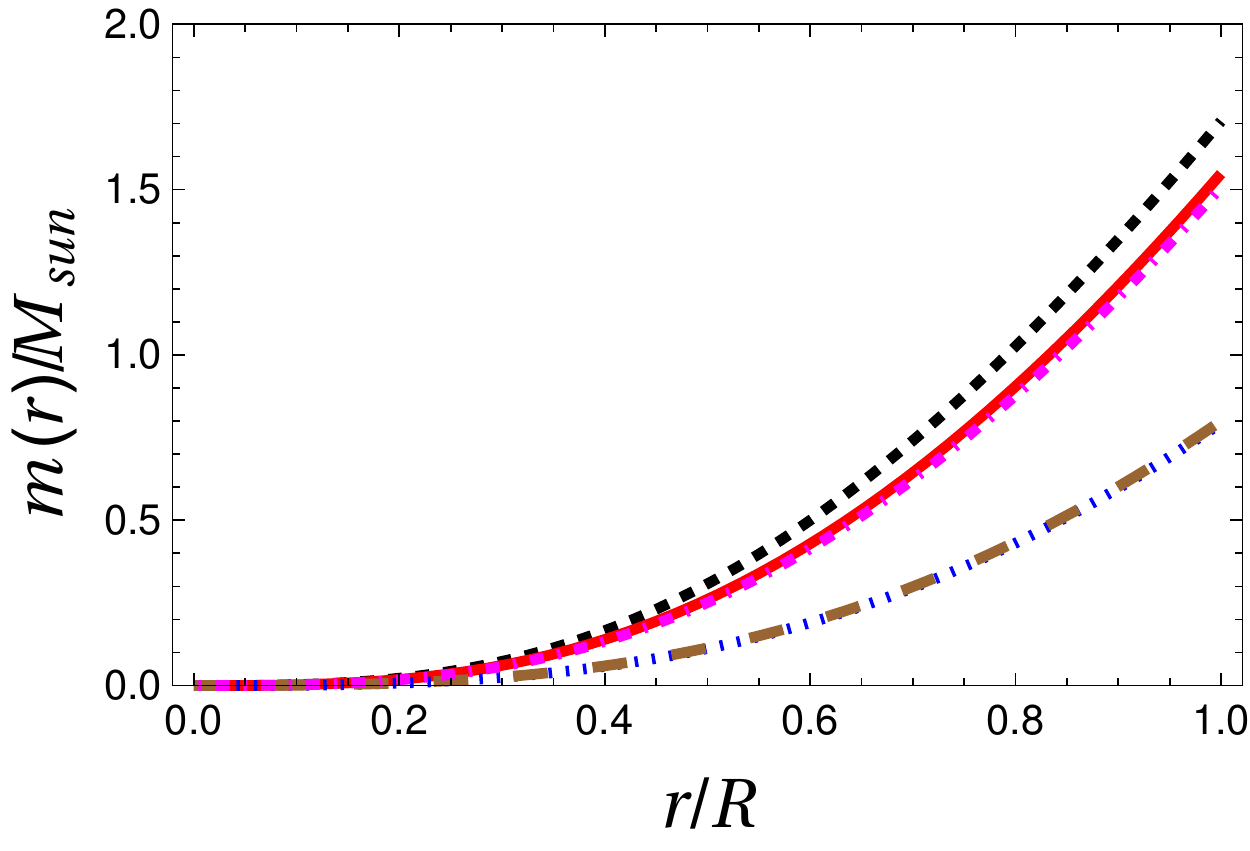}   \
\caption{
Five interior solutions assuming a linear EoS and $B = 60~MeV/fm^3$. Shown are the quantities of interest versus normalized radial coordinate $r/R$. For the initial conditions see Table 1 and text. First case (dashed black line) corresponds to GR.
{\bf{LEFT:}} Gravitation coupling $G(r)$ versus normalized radial coordinate $r/R$ for all five scale-dependent interior solutions obtained here. 
{\bf{MIDDLE:}} Metric potential $\text{e}^{2\nu}$ versus normalized radial coordinate $r/R$.
{\bf{RIGHT:}} Mass function (in solar masses) $m(r)/M_{\text{sun}}$ versus normalized radial coordinate $r/R$. 
}
\label{fig:1}
\end{figure*}



\begin{figure*}[ht]
\centering
\includegraphics[width=0.45\textwidth]{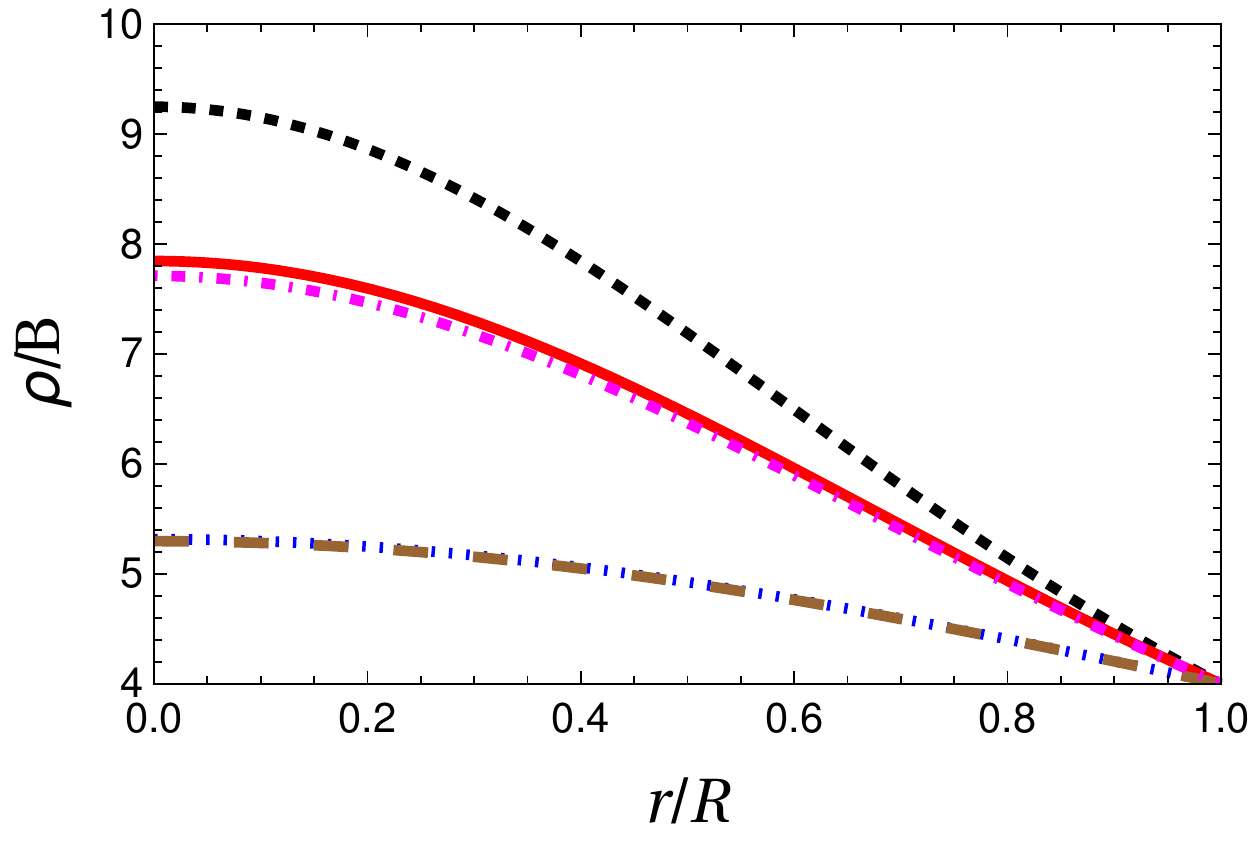}     \
\includegraphics[width=0.45\textwidth]{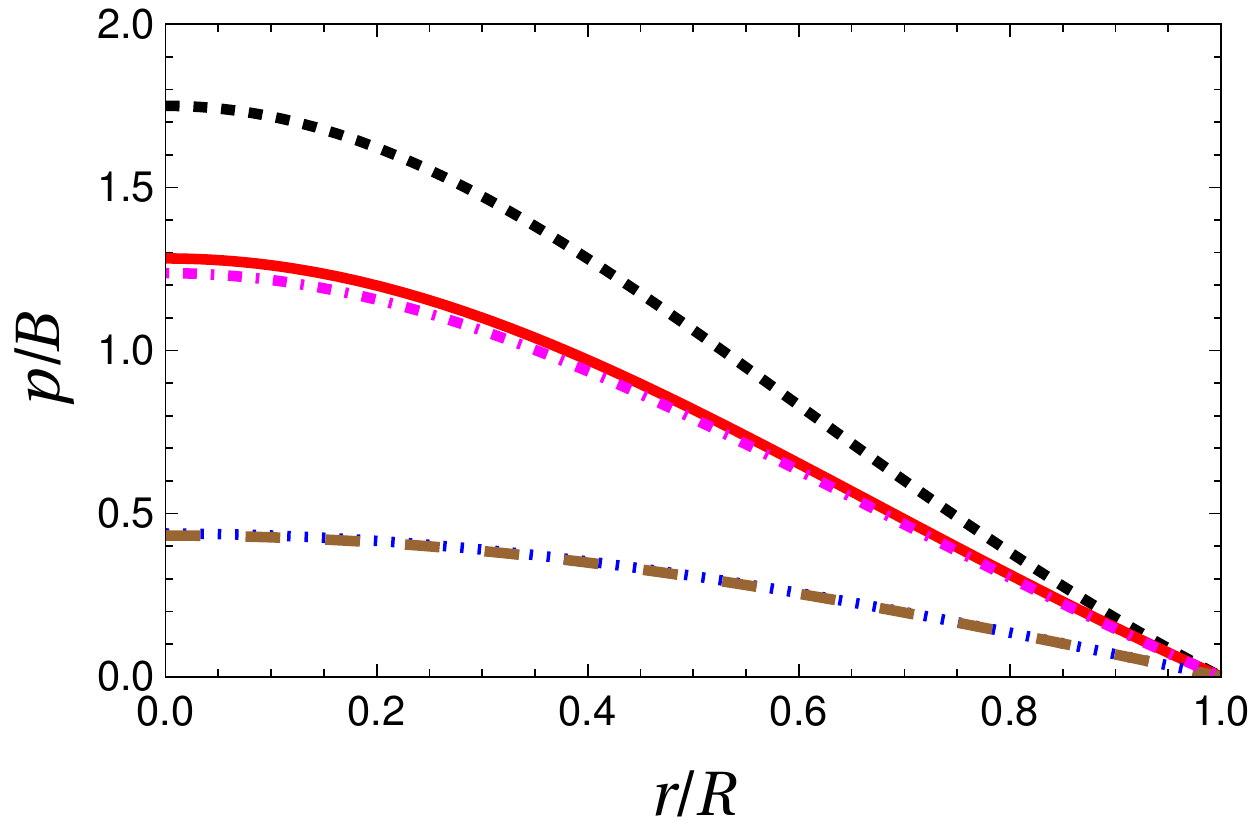}    \
\caption{
Same as before, but for normalized energy density, $\rho/B$, and normalized pressure, $p/B$.
{\bf{LEFT:}} Dimensionless energy density $\rho/B$ versus normalized radial coordinate $r/R$ for all five scale-dependent interior solution obtained in this work. 
{\bf{RIGHT:}} Dimensionless pressure $p/B$ versus normalized radial coordinate $r/R$ for the five scale-dependent interior solutions. 
}
\label{fig:2}
\end{figure*}



\begin{figure*}[ht]
\centering
\includegraphics[width=0.6\textwidth]{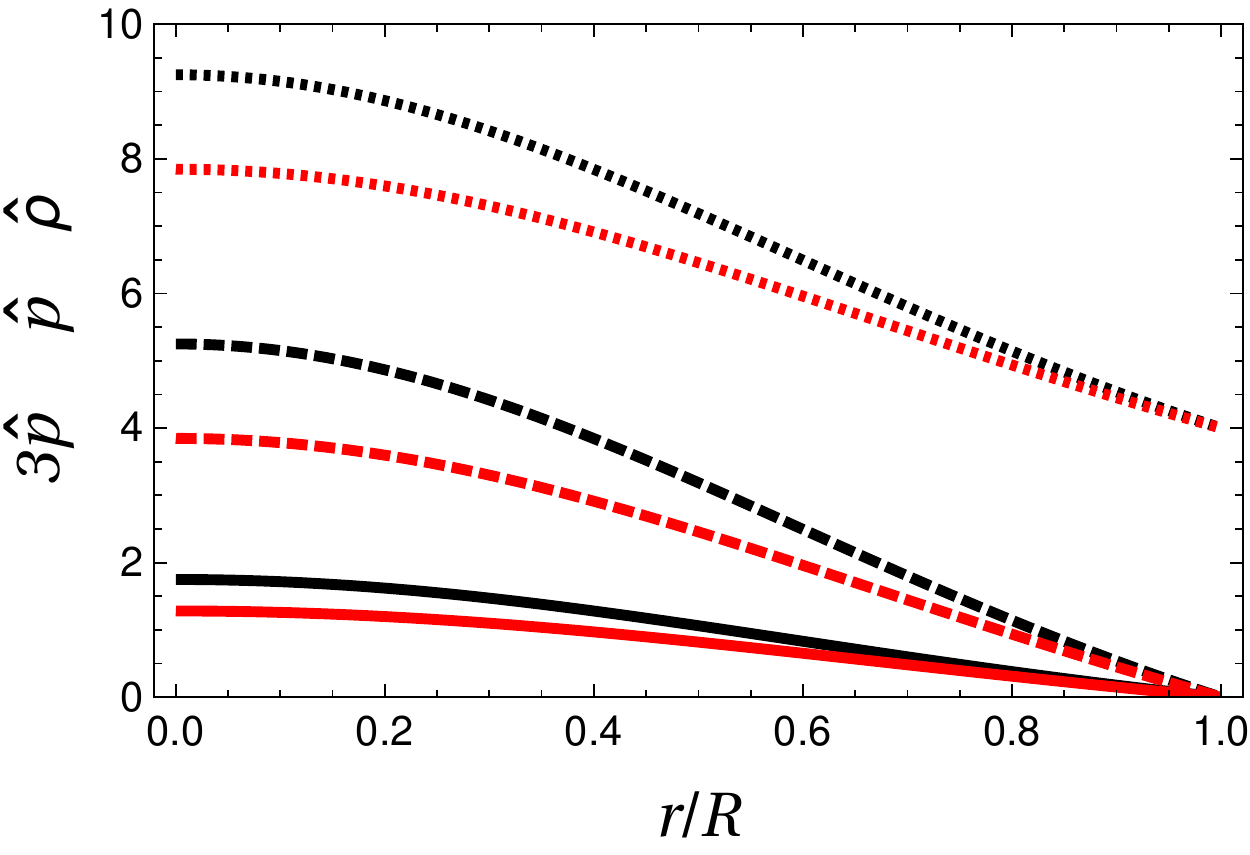}    
\caption{
Energy conditions: Normalized pressure $p/B$ (solid, lower curves), normalized energy density 
$\rho/B$ (dotted, upper curves) and $3p/B$ (dashed curves) versus dimensionless radial coordinate $r/R$ for GR (curves in black) and for the solution $p_c = 1.288~B$ and $\nu_c = -0.475$ (curves in red).
}
\label{fig:3}
\end{figure*}


Our numerical results for the quantities of interest, namely
$p(r),\rho(r),m(r),\nu(r),G(r)$, are summarized in Figures \ref{fig:1} and \ref{fig:2}, and also in Table 1 where we show the mass, the radius and the compactness of the stars as well as the initial conditions for $G(r)$. 

We have obtained five interior solutions assuming different central values for the gravitational coupling, and imposing different central values $p_c$ and $\nu_c$
such that the following conditions are satisfied
\begin{equation}
G(r=R) = 1, \; \; \; \; \; \; \; e^{2 \nu(r=R)}=1-\frac{2M}{R}
\end{equation}
The first case (dashed black line) corresponds to GR, which is also shown for comparison reasons. In particular, the five solutions obtained in this work and shown in Figures \ref{fig:1} and \ref{fig:2} correspond to the following initial conditions: i) $p_c = 1.75~B$ and $\nu_c = -0.557$ (dashed black line), ii) $p_c = 1.288~B$ and $\nu_c = -0.475$ (solid blue line), iii) $p_c = 0.45~B$ and $\nu_c = -0.237$ (dotted blue line), iv) $p_c = 0.44~B$ and $\nu_c = -0.2363$ (long dashed brown line), and finally v) $p_c = 1.245~B$ and $\nu_c = -0.465$. 

Fig.~\ref{fig:1} (left panel) shows the scale-dependent gravitational coupling as a function of radial coordinate. Both positive and negative values of $G(r)$ lead to realistic interior solutions. The horizontal line (dashed black line) corresponds to the case where $G(r)=constant=1$, which is the GR case. All our numerical solutions tend to $G(r=R) = 1$ at the surface of the star, which was imposed right from the start. As it is expected, the deviation from GR is indeed quite small. Fig. \ref{fig:1} (middle panel) depicts the metric potential $e^{2 \nu(r)}$ as a function of the radial coordinate. We notice that the inclusion of a r-varying gravitational coupling shifts the potential to higher values in comparison with the classical counterpart (dashed black line). Finally, in Fig.~ \ref{fig:1} (right) we show the mass function $m(r)$ in solar masses versus normalized radial coordinate. Our results show that the mass function of the compact object is lower than in GR, irrespectively of the sign of the gravitational coupling.

Fig.~\ref{fig:2} (left) shows the energy density versus normalized radial coordinate for the five cases considered here. Once again, we observe that the scale-dependence inclusion induces a non-trivial deviation to the density, which is a physical quantity of the model. Similarly, the pressure is a monotonically decreasing function, lower than the GR prediction. In Table 1 we show the mass $M$, the radius $R$, the compactness $C \equiv M/R$ as well as the initial conditions for the gravitational coupling, $G_c,G_1$. Clearly, the r-varying nature of $G(r)$ leads to smaller and lighter strange quark stars, which are found to be less compact, irrespectively of the sign of $G_1$.

Next, and before we finish, a check regarding the energy conditions should be made. The obtained solutions must be able to describe realistic astrophysical configurations. Therefore, we should investigate if the energy conditions are fulfilled or not. We require that \cite{Ref_Extra_1,Ref_Extra_2,Ref_Extra_3}
\begin{equation}
\rho \geq 0
\end{equation}
\begin{equation}
\rho + p  \geq 0
\end{equation}
\begin{equation}
\rho - p  \geq  0
\end{equation}
\begin{equation}
\rho - 3 p \geq 0
\end{equation}
\begin{equation}
\rho + 3 p \geq 0
\end{equation}
We plot the (normalized) quantities $\rho/B, p/B, 3p/B$ in Fig.~ \ref{fig:3} for GR (in black) and for one of the other four solutions (in red). We observe i) that they are positive and ii) that the energy density remains larger that both $p$ and $3p$ throughout the star. Since all three quantities are positive, clearly the first, the second and the last energy conditions are fulfilled. Moreover, since $\rho$ always remains higher than $p$ and $3p$, the other two energy conditions are fulfilled as well. We have obtained qualitatively similar behaviour for the rest of the solutions. We thus conclude that the interior solutions obtained in the present work are realistic solutions within the framework adopted here, and as such they are able to describe realistic astrophysical configurations.

As a final remark it should be stated here that in the present article we took a first step towards the investigation of spherically symmetric strange quark stars in the scale-dependent scenario assuming the simplest MIT bag model equation-of-state. Clearly, there is still a lot of work to be done. For instance, one may study i) more sophisticated equations-of-state for quark matter, ii) rotating stars, or iii) other compact objects, such as neutron stars or white dwarfs. We hope to be able to address some of those interesting issues in future works.

\section{Conclusions}

Summarizing our work, in the present article we have obtained interior solutions of relativistic stars in the scale-dependence scenario. In particular, we have studied strange quark stars assuming for quark matter the simplest MIT bag model equation-of-state. First we presented the new structure equations describing the hydrostatic equilibrium of the stars. The new equations generalize the usual TOV equations of GR, and they boil down to those when Newton's constant is taken to be a constant, $G'(r)=0=G''(r)$. Then we numerically integrated the structure equations, and we computed the radius and the mass of the stars for a varying Newton's constant, both increasing and decreasing throughout the objects. We also have shown that the energy conditions are fulfilled.
Our numerical results indicate that the r-varying nature of the gravitational coupling in the SD scenario leads to less compact strange quark stars, which are found to be both lighter and smaller, irrespectively of the sign of $G_1$.

\section*{Acknowlegements}

The authors G.~P. and I.~L. thank the Fun\-da\c c\~ao para a Ci\^encia e Tecnologia (FCT), Portugal, for the financial support to the Center for Astrophysics and Gravitation-CENTRA, Instituto Superior T\'ecnico, Universidade de Lisboa, through the Projects No.~UIDB/00099/2020 and No.~PTDC/FIS-AST/28920/2017. The author A.~R. acknowledges DI-VRIEA for financial support through Proyecto Postdoctorado 2019 VRIEA-PUCV.


\begin{thebibliography}{99}

\bibitem{GR} A.~Einstein, 
Annalen Phys. 49 (1916) 769-822.  

\bibitem{tests1} S.~G.~Turyshev,
  Ann.\ Rev.\ Nucl.\ Part.\ Sci.\  {\bf 58} (2008) 207.
  
\bibitem{tests2} C.~M.~Will,
  Living Rev.\ Rel.\  {\bf 17} (2014) 4.
  
\bibitem{tests3} E.~Asmodelle,
  arXiv:1705.04397 [gr-qc].  
  
\bibitem{Ligo} B.~P.~Abbott {\it et al.} [LIGO Scientific and Virgo Collaborations],
  Phys.\ Rev.\ Lett.\  {\bf 116} (2016) no.6,  061102
  [arXiv:1602.03837 [gr-qc]].

\bibitem{QG1} T.~Jacobson,
  Phys.\ Rev.\ Lett.\  {\bf 75} (1995) 1260
  [gr-qc/9504004].

\bibitem{QG2} A.~Connes,
  Commun.\ Math.\ Phys.\  {\bf 182} (1996) 155
  [hep-th/9603053].

\bibitem{QG3} M.~Reuter,
  Phys.\ Rev.\ D {\bf 57} (1998) 971
  [hep-th/9605030].

\bibitem{QG4} C.~Rovelli,
  Living Rev.\ Rel.\  {\bf 1} (1998) 1
  [gr-qc/9710008].

\bibitem{QG5} R.~Gambini and J.~Pullin,
  Phys.\ Rev.\ Lett.\  {\bf 94} (2005) 101302
  [gr-qc/0409057].

\bibitem{QG6} A.~Ashtekar,
  New J.\ Phys.\  {\bf 7} (2005) 198
  [gr-qc/0410054].

\bibitem{QG7} P.~Nicolini,
  Int.\ J.\ Mod.\ Phys.\ A {\bf 24} (2009) 1229
  [arXiv:0807.1939 [hep-th]].

\bibitem{QG8} P.~Horava,
  Phys.\ Rev.\ D {\bf 79} (2009) 084008
  [arXiv:0901.3775 [hep-th]].

\bibitem{QG9} E.~P.~Verlinde,
  JHEP {\bf 1104} (2011) 029
  [arXiv:1001.0785 [hep-th]].
  
\bibitem{SD1} B.~Koch, I.~A.~Reyes and \'A.~Rinc{\'o}n,
  Class.\ Quant.\ Grav.\  {\bf 33} (2016) no.22,  225010
  [arXiv:1606.04123 [hep-th]].  
  
\bibitem{SD2} \'A.~Rinc{\'o}n, E.~Contreras, P.~Bargue{\~n}o, B.~Koch, G.~Panotopoulos and A.~Hern{\'a}ndez-Arboleda,
  Eur.\ Phys.\ J.\ C {\bf 77} (2017) no.7,  494
  [arXiv:1704.04845 [hep-th]].  
  
\bibitem{SD3} \'A.~Rinc{\'o}n and G.~Panotopoulos,
  Phys.\ Rev.\ D {\bf 97} (2018) no.2,  024027
  [arXiv:1801.03248 [hep-th]].

\bibitem{SD4} E.~Contreras, \'A.~Rinc{\'o}n, B.~Koch and P.~Bargue{\~n}o,
  Eur.\ Phys.\ J.\ C {\bf 78} (2018) no.3,  246
  [arXiv:1803.03255 [gr-qc]].

\bibitem{SD5} \'A.~Rinc{\'o}n and B.~Koch,
  Eur.\ Phys.\ J.\ C {\bf 78} (2018) no.12,  1022
  [arXiv:1806.03024 [hep-th]].
  
\bibitem{SD6} \'A.~Rinc{\'o}n, E.~Contreras, P.~Bargue{\~n}o, B.~Koch and G.~Panotopoulos,
  Eur.\ Phys.\ J.\ C {\bf 78} (2018) no.8,  641
  [arXiv:1807.08047 [hep-th]].    

\bibitem{SD7} \'A.~Rinc{\'o}n, E.~Contreras, P.~Bargueño and B.~Koch,
  Eur.\ Phys.\ J.\ Plus {\bf 134} (2019) no.11,  557
  [arXiv:1901.03650 [gr-qc]].
  
\bibitem{ourprd} E.~Contreras, Á.~Rincón, G.~Panotopoulos, P.~Bargueño and B.~Koch,
  Phys.\ Rev.\ D {\bf 101} (2020) no.6,  064053
  [arXiv:1906.06990 [gr-qc]].

\bibitem{alfio} A.~Bonanno, R.~Casadio and A.~Platania,
  JCAP {\bf 2001} (2020) no.01,  022
  [arXiv:1910.11393 [gr-qc]].
  
\bibitem{cand1} J.~A.~Henderson and D.~Page,
  Astrophys.\ Space Sci.\  {\bf 308} (2007) 513
  [astro-ph/0702234 [ASTRO-PH]].
  
\bibitem{cand2} A.~Li, G.~X.~Peng and J.~F.~Lu,
  Res.\ Astron.\ Astrophys.\  {\bf 11} (2011) 482.

\bibitem{cand3} A.~Aziz, S.~Ray, F.~Rahaman, M.~Khlopov and B.~K.~Guha,
  Int.\ J.\ Mod.\ Phys.\ D {\bf 28} (2019) no.13,  1941006
  [arXiv:1906.00063 [gr-qc]].

\bibitem{weber} F.~Weber,
  Prog.\ Part.\ Nucl.\ Phys.\  {\bf 54} (2005) 193.  

\bibitem{Bonanno:2000ep} A.~Bonanno and M.~Reuter,
  Phys.\ Rev.\ D {\bf 62}, 043008 (2000)

\bibitem{Bagnuls:2000ae} C.~Bagnuls and C.~Bervillier,
  Phys.\ Rept.\  {\bf 348}, 91 (2001)

\bibitem{Bonanno:2001xi} A.~Bonanno and M.~Reuter,
  Phys.\ Rev.\ D {\bf 65}, 043508 (2002)

\bibitem{Reuter:2003ca} M.~Reuter and H.~Weyer,
  Phys.\ Rev.\ D {\bf 69}, 104022 (2004)

\bibitem{Koch:2016uso} B.~Koch, I.~A.~Reyes and \'A.~Rinc\'on,
  Class.\ Quant.\ Grav.\  {\bf 33}, no. 22, 225010 (2016)

\bibitem{Rincon:2017ypd} \'A.~Rinc{\'o}n, B.~Koch and I.~Reyes,
  J.\ Phys.\ Conf.\ Ser.\  {\bf 831}, no. 1, 012007 (2017)

\bibitem{Rincon:2017goj} \'A.~Rinc\'on, E.~Contreras, P.~Bargue{\~n}o, B.~Koch, G.~Panotopoulos and A.~Hern{\'a}ndez-Arboleda,
  Eur.\ Phys.\ J.\ C {\bf 77}, no. 7, 494 (2017)
  [arXiv:1704.04845 [hep-th]].

\bibitem{Rincon:2017ayr} \'A.~Rinc\'on and B.~Koch,
  J.\ Phys.\ Conf.\ Ser.\  {\bf 1043}, no. 1, 012015 (2018)
  [arXiv:1705.02729 [hep-th]].
  
\bibitem{Contreras:2017eza} E.~Contreras, \'A.~Rinc\'on, B.~Koch and P.~Bargue{\~n}o,
  Int.\ J.\ Mod.\ Phys.\ D {\bf 27}, no. 03, 1850032 (2017)
  [arXiv:1711.08400 [gr-qc]].

\bibitem{Rincon:2018sgd} \'A.~Rinc\'on and G.~Panotopoulos,
  Phys.\ Rev.\ D {\bf 97}, no. 2, 024027 (2018)
  [arXiv:1801.03248 [hep-th]].

\bibitem{Contreras:2018dhs} E.~Contreras, \'A.~Rin\'on, B.~Koch and P.~Bargue{\~n}o,
  Eur.\ Phys.\ J.\ C {\bf 78}, no. 3, 246 (2018)
  [arXiv:1803.03255 [gr-qc]].  
  
\bibitem{Rincon:2018lyd} \'A.~Rinc\'on and B.~Koch,
  Eur.\ Phys.\ J.\ C {\bf 78}, no. 12, 1022 (2018)
  [arXiv:1806.03024 [hep-th]].

\bibitem{Rincon:2018dsq} \'A.~Rinc\'on, E.~Contreras, P.~Bargue{\~n}o, B.~Koch and G.~Panotopoulos,
  Eur.\ Phys.\ J.\ C {\bf 78}, no. 8, 641 (2018)
  [arXiv:1807.08047 [hep-th]].

\bibitem{Contreras:2018gct} E.~Contreras, \'A.~Rinc\'on and J.~M.~Ramírez-Velasquez,
  Eur.\ Phys.\ J.\ C {\bf 79}, no. 1, 53 (2019)
  [arXiv:1810.07356 [gr-qc]].

\bibitem{Rincon:2019cix} \'A.~Rinc\'on, E.~Contreras, P.~Bargueño and B.~Koch,
  Eur.\ Phys.\ J.\ Plus {\bf 134}, no. 11, 557 (2019)
  [arXiv:1901.03650 [gr-qc]].

\bibitem{Rincon:2019zxk} \'A.~Rinc\'on and J.~R.~Villanueva,
  arXiv:1902.03704 [gr-qc].

\bibitem{Contreras:2019fwu} E.~Contreras, \'A.~Rinc\'on and P.~Bargueño,
  arXiv:1902.05941 [gr-qc].

\bibitem{Fathi:2019jid} 
  M.~Fathi, \'A.~Rinc\'on and J.~R.~Villanueva,
  arXiv:1903.09037 [gr-qc].

\bibitem{Panotopoulos:2019qjk} G.~Panotopoulos and \'A.~Rinc\'on,
  Eur.\ Phys.\ J.\ Plus {\bf 134}, no. 6, 300 (2019)
  [arXiv:1904.10847 [gr-qc]].

\bibitem{Contreras:2019nih} E.~Contreras, J.~M.~Ramirez-Velasquez, \'A.~Rinc{\'o}n, G.~Panotopoulos and P.~Bargueño,
  Eur.\ Phys.\ J.\ C {\bf 79}, no. 9, 802 (2019)
  [arXiv:1905.11443 [gr-qc]].

\bibitem{Hernandez-Arboleda:2018qdo} A.~Hernández-Arboleda, \'A.~Rinc\'on, B.~Koch, E.~Contreras and P.~Bargueño,
  arXiv:1802.05288 [gr-qc].

\bibitem{Canales:2018tbn} 
  F.~Canales, B.~Koch, C.~Laporte and \'A.~Rinc\'on,
  JCAP {\bf 2001}, no. 01, 021 (2020)
  [arXiv:1812.10526 [gr-qc]].
  
\bibitem{Koch:2010nn} B.~Koch and I.~Ramirez,
  Class.\ Quant.\ Grav.\  {\bf 28}, 055008 (2011)
  [arXiv:1010.2799 [gr-qc]].     
  
\bibitem{Tolman} R.~C.~Tolman,
  Phys.\ Rev.\  {\bf 55}, 364 (1939).
   
\bibitem{OV} J.~R.~Oppenheimer and G.~M.~Volkoff,
Phys.\ Rev.\  {\bf 55} (1939) 374.  

\bibitem{SBH} K.~Schwarzschild,
Sitzungsber.\ Preuss.\ Akad.\ Wiss.\ Berlin (Math.\ Phys.\ ) {\bf 1916} 
(1916) 189 [physics/9905030]. 

\bibitem{formalism} \'A.~Rinc{\'o}n and B.~Koch,
  J.\ Phys.\ Conf.\ Ser.\  {\bf 1043} (2018) no.1, 012015
  [arXiv:1705.02729 [hep-th]].
  
\bibitem{bagmodel1} A.~Chodos, R.~L.~Jaffe, K.~Johnson, C.~B.~Thorn and V.~F.~Weisskopf,
  Phys.\ Rev.\ D {\bf 9} (1974) 3471.

\bibitem{bagmodel2} A.~Chodos, R.~L.~Jaffe, K.~Johnson and C.~B.~Thorn,
  Phys.\ Rev.\ D {\bf 10} (1974) 2599. 
  
\bibitem{farhi} E.~Farhi and R.~L.~Jaffe,
  Phys.\ Rev.\ D {\bf 30} (1984) 2379.
  
\bibitem{simpleMIT} D.~Gondek-Rosinska and F.~Limousin,
  arXiv:0801.4829 [gr-qc].  
  
\bibitem{wilczek1} M.~G.~Alford, K.~Rajagopal and F.~Wilczek,
  Phys.\ Lett.\ B {\bf 422} (1998) 247
  [hep-ph/9711395].

\bibitem{wilczek2} M.~G.~Alford, K.~Rajagopal and F.~Wilczek,
  Nucl.\ Phys.\ A {\bf 638} (1998) 515C
  [hep-ph/9802284].

\bibitem{wilczek3} M.~G.~Alford, K.~Rajagopal and F.~Wilczek,
  Nucl.\ Phys.\ B {\bf 537} (1999) 443
  [hep-ph/9804403].

\bibitem{wilczek4} K.~Rajagopal and F.~Wilczek,
  Phys.\ Rev.\ Lett.\  {\bf 86} (2001) 3492
  [hep-ph/0012039].

\bibitem{NJL1} Y.~Nambu and G.~Jona-Lasinio,
  Phys.\ Rev.\  {\bf 122} (1961) 345.

\bibitem{NJL2} Y.~Nambu and G.~Jona-Lasinio,
  Phys.\ Rev.\  {\bf 124} (1961) 246.

\bibitem{refine1} E.~S.~Fraga, A.~Kurkela and A.~Vuorinen,
  Astrophys.\ J.\  {\bf 781} (2014) no.2,  L25
  [arXiv:1311.5154 [nucl-th]].
  
\bibitem{art} A.~Vuorinen,
  arXiv:1611.04557 [hep-ph].  

\bibitem{refine2} V.~D.~Toneev, E.~G.~Nikonov, B.~Friman, W.~Norenberg and K.~Redlich,
  Eur.\ Phys.\ J.\ C {\bf 32} (2003) 399
  [hep-ph/0308088].
  
\bibitem{refine3} Y.~B.~Ivanov, A.~S.~Khvorostukhin, E.~E.~Kolomeitsev, V.~V.~Skokov, V.~D.~Toneev and D.~N.~Voskresensky,
  Phys.\ Rev.\ C {\bf 72} (2005) 025804
  [astro-ph/0501254].
  
\bibitem{Ref_Extra_1} M.~K. Mak and T.~Harko, Chin. J. J. Astron. Astrophys, {\bf 2}, 248 (2002).

\bibitem{Ref_Extra_2} D.~Deb, S.~R.~Chowdhury, S.~Ray, F.~Rahaman and B.~K.~Guha,
  Annals Phys.\  {\bf 387}, 239 (2017)

\bibitem{Ref_Extra_3} D.~Deb, S.~Roy Chowdhury, S.~Ray and F.~Rahaman,
  Gen.\ Rel.\ Grav.\  {\bf 50}, no. 9, 112 (2018).
  
\end{thebibliography}
\end{document}